\documentclass[prl,twocolumn,aps,superscriptaddress,showpacs,floatfix]{revtex4}
\usepackage{graphicx}
\usepackage{colordvi}
\usepackage{dcolumn}
\usepackage{amsmath}
\usepackage{slashbox}

\begin{document}

\title{Wavelength dependence of high-order harmonic generation with the ionization and ponderomotive energy controlled by an extreme ultraviolet pulse}

\author{Kenichi L. Ishikawa}
\email[Electronic address: ]{ishiken@riken.jp}
\affiliation{Integrated Simulation of Living Matter Group, RIKEN Computational Science Research Program, 2-1 Hirosawa, Wako, Saitama 351-0198, Japan}
\affiliation{PRESTO (Precursory Research for Embryonic Science and Technology), Japan Science and Technology Agency, Honcho 4-1-8, Kawaguchi-shi, Saitama 332-0012, Japan}

\author{Eiji J. Takahashi}
\affiliation{Extreme Photonics Research Group, 
RIKEN Advanced Science Institute, 2-1 Hirosawa, Wako, Saitama 351-0198, Japan }

\author{Katsumi Midorikawa}
\affiliation{Extreme Photonics Research Group, RIKEN Advanced Science Institute, 2-1 Hirosawa, Wako, Saitama 351-0198, Japan }

\date{\today}

\begin{abstract} 
We theoretically study the scaling with the driving wavelength $\lambda$ of the high-order harmonic generation (HHG) under the simultaneous irradiation of an extreme ultraviolet (XUV) pulse. Surprisingly, when the cutoff energy and ionization yield are fixed, the harmonic yield is nearly independent of $\lambda$. We identify its origin as the combination of the initial spatial width of the states excited by the XUV pulse, making the wavepacket spreading less prominent, and the shallowing of the ionization potential, which suggests complex nature of the wavelength dependence of HHG.
\end{abstract}

\maketitle

High-order harmonic generation (HHG) represents one of the best methods to produce ultrashort coherent
light covering a wavelength range from the vacuum ultraviolet to the soft X-ray region.
The maximal harmonic photon energy $E_c$ is given by the cutoff law $E_c = I_p + 3.17 U_p$ \cite{Krause1992PRL}, 
where $I_p$ is the ionization potential of the target atom, and $U_p  {\rm [eV]}= F^2/4\omega^2=9.337 \times 10^{-14}~I~{\rm [W/cm^2]} ~(\lambda~{\rm [\mu m]})^2$ the ponderomotive energy, with $F$, $I$ and $\lambda$ being the strength, intensity and wavelength of the driving field, respectively. Since $U_p$ scales as $\lambda^2$, a promising route to generate harmonics of higher photon energy is to use a driving laser of a longer wavelength.
Thus, the laser wavelength ($\lambda$) is an effective control knob for the ponderomotive energy, and hence the cutoff.
This has motivated HHG experiments with high-power mid-infrared (MIR) lasers \cite{NatPhys.4.386.2008,tak_CLEO2008,Popmintchev:08}. Using a 1.55 $\mu$m driving laser field from an optical parametric amplifier \cite{tak_CLEO2008}, for example, Takahashi $et ~al.$ \cite{tak_CLEO2008_2} have recently succeeded in generating harmonics with a photon energy of 300 eV from Ne and 450 eV from He gas, which lie well in the water-window region.

Under such a circumstance, the dependence of the HHG yield on $\lambda$ has become an issue of increasing interest.
Although it had been commonly assumed that the HHG efficiency scaled as $\lambda^{-3}$ due to the spreading of the returning wavepacket \cite{Lewenstein1994PRA},
recent theoretical\cite{tate_013901,Schiessl2007PRL,Schiessl2008JMO,Ishikawa2009PRA,Frolov_PRL} as well as experimental\cite{NatPhys.4.386.2008,tak_CLEO2008} studies have revealed much stronger dependence of $\propto \lambda^{-x}$ with $5 \le x \le 6$, which would significantly reduce the HHG yield by MIR lasers.
It is considered that the additional factor $\lambda^{-2}$ is of an apparent nature stemming from the distribution of the HHG energy up to the cutoff ($\propto\lambda^2$) \cite{Schiessl2007PRL,Schiessl2008JMO}, though the precise physical origin of the scaling law has not been fully understood yet.

\begin{figure} 
\centerline{\includegraphics[width=8.3cm]{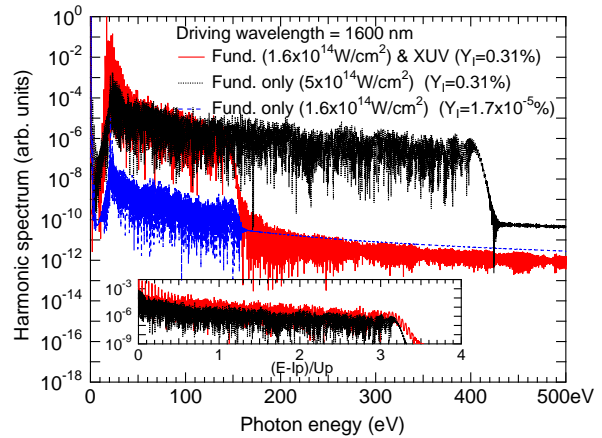}} 
\caption{\label{fig1} (color online) Upper solid curve: harmonic spectrum from ${\rm He}$ exposed to a 35 fs Gaussian combined driving and XUV pulse ($\hbar\omega_X=17.05\,{\rm eV}$), the former ($\lambda=1600\ {\rm nm}$) with a peak intensity of $1.6\times 10^{14}\,{\rm W/cm}^2$ and the latter $2.3\times 10^{11}\,{\rm W/cm}^2$. Middle dotted and lower dashed curves: harmonic spectra for the cases of the driving pulse alone, with an intensity of $5\times 10^{14}\,{\rm W/cm}^2$ and $1.6\times 10^{14}\,{\rm W/cm}^2$, respectively. Inset: replots of the upper two curves in terms of $(\hbar\omega_h-I_p)/U_p$.}
\end{figure}

While most of the experiments are conducted with a driving laser of a single wavelength, the control of HHG using XUV pulses has also been discussed \cite{Schafer2004PRL,Ishikawa2003PRL,Ishikawa2004PRA}. For example, Schafer {\it et al.} \cite{Schafer2004PRL} showed that the delay of attosecond pulse trains can be used to microscopically select a single quantum path contribution.
On the other hand, Ishikawa \cite{Ishikawa2003PRL,Ishikawa2004PRA} has theoretically shown that the irradiation of the XUV pulse with a photon energy $\hbar\omega_{X}$ smaller than $I_p$ can boost the ionization $Y_I$ and harmonic yield $Y_H$ by orders of magnitude; the XUV pulse facilitates optical-field ionization by promoting a transition to (real or virtual) excited states. This effect has been experimentally demonstrated by the use of mixed gases \cite{Takahashi2006UP}, and its application to single attosecond pulse generation has been proposed \cite{AEGIS2007PRA}. One of the remarkable features of this effect is that $Y_I$ increases in proportion to the XUV intensity \emph{without affecting the cutoff energy determined by the driving infrared pulse} (Fig.\ \ref{fig1}). Thus, the addition of XUV pulses can be viewed as a tool to enable independent control of $\lambda$, $E_c$, and $Y_I$; for a given value of $\lambda$, $E_c$ can be adjusted through $I$ and then $Y_I$ through the XUV intensity. This would provide the investigation of the $\lambda$-dependence of the HHG with additional degrees of freedom. 

The above consideration has motivated us to theoretically investigate the driving-wavelength-dependence of HHG with the XUV control of $E_c (U_p)$ and $Y_I$. For the case of the driving laser pulse alone, if we fix $E_c$ at each driving wavelength, the driving intensity is lowered with an increasing wavelength, leading to the drop of $Y_I$, which in turn largely affects the HHG efficiency. The addition of an XUV pulse of appropriate intensity, however, can adjust $Y$ to a constant value, and, then, we would expect $\propto \lambda^{-3}$ scaling due to the wavepacket spreading. Our results based on numerical solution of the time-dependent Schr\"odinger equation (TDSE), however, show that the harmonic yield is nearly independent of $\lambda$ at fixed ponderomotive energy and ionization. Using the Lewenstein model \cite{Lewenstein1994PRA}, we identify the origin of this surprising feature as the combination of the initial spatial width of the wave function and shallowing of the effective ionization potential, indicating complex nature of the $\lambda$-dependence of HHG.

To study the single-atom response under a combined driving laser and XUV pulse, we solve the TDSE in the length gauge,
\begin{equation}
    i\frac{\partial\psi({\bf r},t)}{\partial t} =
     \left[-\frac{1}{2}\nabla^2+V(r)+z[E(t)+E_X(t)]\right]\psi({\bf r},t),
\label{eq:01}
\end{equation}
for a model atom in the single active electron approximation, represented by an effective potential \cite{Muller1999PRA},
\begin{equation}
V(r) =  -[1+\alpha e^{-r}+(Z-1-\alpha)e^{-\beta r}]/r,
\label{eq:02}
\end{equation}
where $Z$ denotes the atomic number. For He, we use parameters $Z=2$, $\alpha=0$, and $\beta=2.157$, which faithfully reproduce the eigenenergies of the ground and the first excited states. $E(t)=Ff(t)\sin\omega t$ is the driving optical field, with $F$ being the peak amplitude and $f(t)$ the envelope function corresponding the Gaussian profile with a full width at half maximum (FWHM) of 35 fs. $E_X(t)=Ff_X(t)\sin\omega_X t$ is the XUV field, with $F_X$ being the peak amplitude. The harmonic spectrum is calculated by Fourier transforming the dipole acceleration, and the HHG yield is defined as energy radiated from the target atom per unit time \cite{Jackson} integrated for a fixed range of photon energy $\hbar\omega_h$, specifically from 30 to 60 eV. 

Figure \ref{fig1} shows the harmonic spectra from He for $\lambda=1600\,{\rm nm}$ with and without the XUV field ($\hbar\omega_X=17.05\,{\rm eV}$). For the case of the driving laser alone with a peak intensity $I$ of $1.6\times 10^{14}\,{\rm W/cm}^2$ (blue dashed curve), the ionization yield $Y_I$ is very low ($1.7\times 10^{-5}\%$). We can increase $Y_I$ in two ways. First, if we augment $I$ to $5\times 10^{14}\,{\rm W/cm}^2$ (black dotted curve), $Y_I$ reaches $0.31\%$ and, accordingly, the harmonic yield becomes higher, which is accompanied by the increase of the cutoff energy. 
Alternatively, the same ionization yield can be achieved by the addition of the XUV pulse with an appropriate intensity of $2.3\times 10^{11}\,{\rm W/cm}^2$. In this case (red solid line), the cutoff remains nearly unchanged. Hence, as already mentioned, the combination of the laser and XUV pulses can be used as a tool to adjust $\lambda$, $E_c (U_p)$, and $Y$ independently.
It should also be noted that the resulting harmonics have an even higher yield than those from a driving laser of higher intensity alone (black dotted line) between 30 and 60 eV. The ratio between the two cases in this energy range is $\approx 3.2$, which is comparable with the ratio of $U_p$. In addition,  as is shown in the inset, the harmonic yield is distributed in a similar manner between $\hbar\omega_h=I_p$ and $E_c$, in spite of the large difference in driving intensity and cutoff energy. These observations are consistent with the idea that the additional scaling $\propto \lambda^{-2}$ is an apparent effect due to the harmonic energy distribution up to the cutoff.

\begin{figure} 
\centerline{\includegraphics[width=8.3cm]{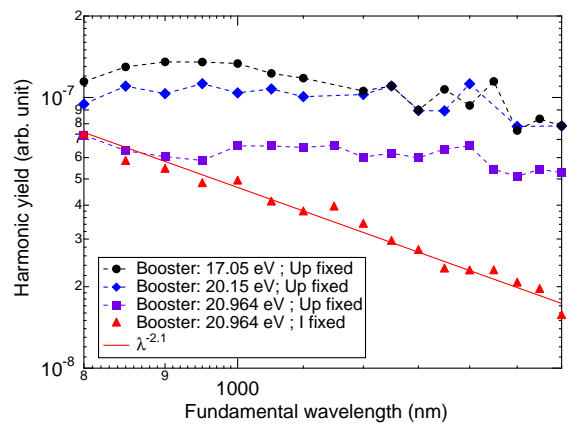}} 
\caption{\label{fig2} (Color online) Wavelength dependence of the TDSE-calculated DE harmonic yield from He between 30 and 60 eV, for different values of $\hbar\omega_X$. $I=1.6\times 10^{14} \times ((800 \ {\rm nm})/\lambda)^2\ {\rm W/cm}^2$, so that $U_p$ may remain unchanged, except for the triangles and the fitting line, for which $I$ is fixed at  $1.6\times 10^{14}\ {\rm W/cm}^2$. The XUV intensity is adjusted in such a way that the ionization yield is 1\%, irrespective of $\lambda$.}
\end{figure}

Encouraged by these results, let us now explore how the harmonic yield variates with the driving wavelength when $U_p (\propto I\lambda^2)$ and ionization are kept constant simultaneously by the addition of an XUV pulse. Many features of HHG can be intuitively and even quantitatively explained by the semi-classical three-step model \cite{Krause1992PRL,Schafer1993PRL,Corkum1993}. According to this model, an electron is lifted to the continuum at the nuclear position with no kinetic energy (\emph{ionization}), the subsequent motion is governed classically by an oscillating electric field (\emph{propagation}), and a harmonic is emitted upon \emph{recombination}. The last step is independent of $\lambda$ as far as a given harmonic photon energy range is concerned. The first step is fixed. Concerning the propagation step, if we neglect $I_p$ in the saddle-point equations \cite{Lewenstein1994PRA,Milosevic2002PRA}, or equivalently, if we consider a classical motion of electron in an oscillating electric field starting from the origin with a vanishing initial velocity, the phases of the field upon ionization $\phi_i=\omega t_i$ and recombination $\phi_r=\omega t_r$ ($t_i,t_r$: time of ionization and recombination, respectively), characterizing quantum trajectories, are a function of $\hbar\omega_h/U_p$, hence common for any value of $\lambda$, since $U_p$ is fixed. Thus, we might expect that the comparison under the condition of fixed ionization and $U_p$ extracts the effect of the wavepacket spreading.
 
In Fig.\ \ref{fig2} we show the dependence of the XUV-assisted harmonic yield on the driving wavelength from 800 nm to 1.6\ $\mu{\rm m}$ for several different values of XUV photon energy $\hbar\omega_X$, including $20.964\ {\rm eV}$ resonant with the transition to the first excited state.
The peak intensity $I$ is $1.6\times 10^{14}\ {\rm W/cm}^2$ at $\lambda=800\ {\rm nm}$ and varied so that $U_p$ ($\propto I\lambda^2$) remains unchanged. The XUV intensity is adjusted to yield $Y=1\%$, irrespective of $\lambda$. We can see that, apart from fluctuations due to quantum-path interference \cite{Schiessl2007PRL,Schiessl2008JMO,Ishikawa2009PRA,Frolov_PRL,Milosevic2008JMO}, the harmonic yield is nearly independent of driving wavelength, in great contrast to the common anticipation that the wavepacket spreading has a contribution $\propto\lambda^{-3}$. In this figure is also shown the result for the driving intensity fixed at $1.6\times 10^{14}\ {\rm W/cm}^2$; the ionization yield is adjusted again to 1\% through the XUV intensity, though it scarcely depends on $\lambda$. In this case, reflecting the apparent harmonic energy distribution effect, the HHG yield scales as $\lambda^{-2}$, which is much gentler than the usual $\lambda^{-5}$ dependence for the case of the driving pulse alone.

In order to clarify the origin of this surprising feature, let us re-examine the wavepacket spreading during the propagation process. The enhancement mechanisms under simultaneous irradiation of the XUV pulse are harmonic generation from a coherent superposition of states and two-color frequency mixing (tunneling ionization from a virtual excited state) \cite{Ishikawa2003PRL,Ishikawa2004PRA}. The excited states are spatially much more extended than the ground state. Our discussion so far as well as the common discussion on the wavelength dependence, however, neglects the initial spatial width of the wave function. The latter can be explicitly accounted for in the Lewenstein model \cite{Lewenstein1994PRA} if we approximate the ground state by a Gaussian wave function $\psi ({\bf r}) = (\pi\Delta^2)^{-3/4} e^{-{\bf r}^2/(2\Delta^2)}$,
where $\Delta\ (\sim I_p^{-1})$ is the spatial width. An appealing point of this Gaussian model is that one can analytically evaluate the integral with respect to momentum in the formula for the dipole moment (Eq.\ (8) of Ref.\ \cite{Lewenstein1994PRA}). The spreading factor $(2\Delta^2+i\tau)^{-3/2}$ in the resulting formula (Eq.\ (22) of Ref.\ \cite{Lewenstein1994PRA}) includes the effect of the width of the initial state. 

Let us here extend the above discussion to the HHG from the superposition of the ground and an excited states, relevant to the enhancement mechanism \cite{Ishikawa2003PRL,Ishikawa2004PRA}. Then, following Ref.\ \cite{Watson1996PRA}, we obtain the formula for the dipole moment $d(t)$ as,
\begin{widetext}
\begin{eqnarray}
\label{eq:SuperpositionGaussian}
d(t) = i(\Delta_g\Delta_e)^{-7/2}\int_{-\infty}^t (2C(t,t^\prime))^{3/2}E(t^\prime)\{A(t)A(t^\prime)+C(t,t^\prime)[1-D(t,t^\prime)(A(t)+A(t^\prime))]+C^2(t,t^\prime)D^2(t,t^\prime)\}\nonumber\\
\times \exp\left(-i[(I_pt-I_et^\prime)+B(t,t^\prime)]
-\frac{A^2(t)\Delta_g^2+A^2(t^\prime)\Delta_e^2-C(t,t^\prime)D^2(t,t^\prime)}{2}\right),
\end{eqnarray}
\end{widetext}
where $I_e$ denotes the ionization potential of the excited level, $\Delta_g$ and $\Delta_e$ the spatial width of the ground and excited states, respectively, $A(t)$ the vector potential, and 
\begin{eqnarray}
B(t,t^\prime)=\frac{1}{2}\int_{t^\prime}^{t}dt^{\prime\prime}A^2(t^{\prime\prime}),\\
C(t,t^\prime)=\left(\Delta_g^2+\Delta_e^2+i(t-t^\prime)\right)^{-1},\\
D(t,t^\prime)=A(t)\Delta_g^2+A(t^\prime)\Delta_e^2+i\int_{t^\prime}^{t}dt^{\prime\prime}A(t^{\prime\prime}).
\end{eqnarray}
The factor $C^{3/2}(t,t^\prime)$ describes the leading contribution from the wavepacket spreading. For the first excited state of He ($I_e=3.6\ {\rm eV}$), for  example, $\Delta_e^2$ is several tens of a.u., hence comparable with the excursion time $\tau=t-t^\prime$. This, making the wavepacket spreading relatively less prominent, is expected to influence the wavelength scaling.

It should be noted that if we resorted to the saddle-point analysis (SPA) \cite{Lewenstein1994PRA,Milosevic2002PRA} instead of the momentum integration, the ionization time $t^\prime$ would contain an imaginary part ${\rm Im}\, t^\prime\approx{\sqrt{2I_{e}}}/{E(t^\prime)}$ stemming from the tunneling process. Then the spreading factor would rather read as $\left(\Delta_g^2+\Delta_e^2+{\sqrt{2I_{e}}}/{E(t^\prime)}+i\tau\right)^{-3/2}$, containing an additional term that can be interpreted as the width at the tunnel exit \cite{Gottlieb1996PRA}. This tunneling contribution is automatically accounted for in Eq.\ (\ref{eq:SuperpositionGaussian}). Since $\Delta_g^2<{\sqrt{2I_{e}}}/{E(t^\prime)}<{\sqrt{2I_{p}}}/{E(t^\prime)}<\Delta_e^2$ in general, the width of the excited state has the largest contribution in the XUV-assisted HHG while the initial width is negligible for the case of the ground-state atom. 

The form of Eq.\,(\ref{eq:SuperpositionGaussian}) suggests that the dependence of the HHG yield on the initial spatial width of the wave function and the ionization potential is rather complex. While these two are correlated to each other in the real atom, here we treat them as independent parameters and list in Table \ref{table:GaussianModel} the exponent of the power-law scaling for different combinations of $\Delta_e$ and $I_e$, calculated with Eq.\ (\ref{eq:SuperpositionGaussian}). $I_e$ as a free parameter may be interpreted as the effective ionization potential defined by $I_e=I_p-\hbar\omega_X$. It should be noted that the peak intensity is fixed at $1.6\times 10^{14}\ {\rm W/cm}^2$, so the results are to be compared with the triangles in Fig.\ \ref{fig2}. Both larger initial spatial width and shallower effective ionization potential decrease the exponent, and their synergy leads to the surprising gentle wavelength scaling.

\begin{table}
\caption{Exponent $x$ of the wavelength scaling $\propto\lambda^{-x}$ for various combinations of the initial spatial width $\Delta_e$ and the effective ionization potential $I_e$.}
\begin{center}
\begin{tabular}{| l||c|c|c|}\hline
\backslashbox{$\Delta_e$ (a.u.)}{$I_e$ (eV)} & 3.6 & 4.4 & 13.6 \\\hline\hline
5.8 & 2.2 & 2.2 & 4.3\\\hline
4.5 & 2.4 & 2.7 & 4.5\\\hline
3.2 & 2.7 & 3.4 & 4.9\\\hline
1.1 & 4.0 & 4.2 & 5.4\\\hline
\end{tabular}
\end{center}
\label{table:GaussianModel}
\end{table}

In summary, we have investigated the driving-wavelength dependence of HHG under the simultaneous irradiation of a non-ionizing XUV pulse. The XUV pulse serves as a tool to provide additional degrees of freedom to the study of the $\lambda$-dependence of HHG, with its ability to adjust the cutoff and ionization yield independently and control the initial spatial width of the wave function and the effective ionization potential. We have shown that the XUV-assisted harmonic yield scales with $\lambda$ much more weakly than for the case of the driving laser alone; fixed $U_p$ and $Y$, especially, lead to a very small $\lambda$ dependence. According to our analysis based on the Gaussian model, the combination of the large spatial width of the states excited by the XUV pulse making the effect of the wavepacket spreading less prominent and the shallowing of the effective ionization potential is responsible for this unexpected feature. While both effects are described in Eq.\ (\ref{eq:SuperpositionGaussian}) in principle, clear-cut explanation why the latter contributes to the gentle scaling is not at hand. The results of the present study indicate that the $\lambda$-scaling of HHG is not simply governed by the wavepacket spreading ($\propto\lambda^{-3}$) and the apparent energy distribution effect ($\propto\lambda^{-2}$), but exhibits richer and more complex behavior than previously considered. There are indeed further open questions such as why higher-order returning trajectories have so important contribution \cite{tate_013901,Schiessl2007PRL,Schiessl2008JMO,Ishikawa2009PRA} and whether the wavepacket spreading should give a factor $\propto\lambda^{-2}$ rather than $\propto\lambda^{-3}$ since the spreading in the direction of the quiver motion is swept over the parant ion upon recollision. Further study will be necessary to answer these questions.

K.L.I. acknowledges inspiring discussions with H. Suzuura, J. Burgd\"orfer, and K. Schiessl. K.L.I. also gratefully acknowledges financial support by the Precursory Research for Embryonic Science and Technology (PRESTO) program of the Japan Science and Technology Agency (JST) and by the Ministry of Education, Culture, Sports, Science, and Technology of Japan, Grant No. 19686006.
This study was financially  supported by a grant from the Research Foundation for Opto-Science and Technology.

\end{document}